\documentclass{ewic}
\input{psfig.txt}

\begin{document}

\header{Spectroscopic surveys:a different approach to data reduction}

\footer{\emph{Astronomical Data Analysis Conference III}\\
}

\title{Spectroscopic surveys: a different approach to data reduction}

\author{B.Garilli, P.Franzetti, L.Paioro, M.Scodeggio, A.Zanichelli\\
INAF-IASF, Milan\\
via Bassini, 15 20133 Milan, Italy\\
\email{bianca@mi.iasf.cnr.it}}

\begin{abstract}
We present VIPGI, an automatized human supervised reduction
environment, developed within the VIRMOS project to handle VIMOS guaranteed
time data. VIPGI is now offered to the international community to be
used on site in Milano and Marseille. Born to handle the highly
multiplexed MOS VIMOS data, it has been extended to accomodate also IFU data. 
The widespread and extensive use of VIPGI has suggested the idea of
an integrated environment allowing users not only to reduce, 
but also to organize data in logical structures, to insert results in a
database, and use any user defined plugin for data reduction, 
analysis and inspection. See http://cosmos.mi.iasf.cnr.it/pandora
\end{abstract}


\section{INTRODUCTION}
In the last years the number of large telescopes available to the 
astronomical community has rapidly increased, 
together with the multiplexing capabilities of the instruments
attached to those telescopes.
While a normal long-slit spectrograph on a 4-meter class telescope 
could produce a few tens of spectra per night of 
observation, today a last generation optical spectrograph can obtain 
several thousands spectra per night. This productivity 
increase has rendered obsolete traditional methods of data reduction
and analysis, at least as long as these data must be 
reduced and analyzed in a timely fashion.
Moreover it is necessary to develop 
an efficient and rigorous data organizer and archiver, so that 
the available data and files would not be lost
among hundreds or thousands of similar data and files.
General-purpose astronomical software packages like IRAF or MIDAS 
are not well equipped for these tasks. They offer a huge 
amount of tasks to manipulate astronomical data which constitute 
the elementary particles with which, in principle, it 
could be possible to build up a semi-automatic pipeline. As a matter 
of facts, however, each step of the reduction process 
requires the execution of too many such ``particles'' to allow for a
substantial automation.
Moreover, such tools do not offer any facility for organizing data, 
an aspect which must not be neglected: dealing 
manually with hundreds of spectra per exposure can be a hard job even 
for the most systematic astronomer.\\

Among the present day spectrographs, VIMOS is perhaps the most
challenging in terms of data production.
VIMOS (\cite{lefevre2003}) is the imaging spectrograph mounted on Unit 3  
(Melipal) of the Very Large Telescope (VLT) at the 
Paranal Observatory.
It has been specially designed to be a survey oriented instrument, 
therefore its multiplexing capabilities have been pushed 
to the maximum: during a single exposure, up to 800 spectra can 
be obtained in MOS mode (6400 in IFU mode), and this means 
a really huge amount of data. 
Difficulties begin already when trying to find one's way in
the jungle of raw science and calibration frames, as distributed via
the standard ESO archive procedures, and increase going further along
the reduction process, due to the number of calibrations and
corrections that are to be applied to the data (at least for
spectroscopic ones). 

For this reason the VIRMOS Consortium was asked
to deliver to ESO all the elements necessary to build an automatic
data reduction pipeline. The result of this work is the VIMOS Data 
Reduction Software (DRS). On top of this, to allow the usage of
the DRS outside the ESO Data Flow System, we have developed VIPGI
(Vimos Interactive Pipeline and Graphical Interface). Based on DRS
for the reduction part, VIPGI provides the astronomer with 
a user-friendly graphical interface for executing the pipeline
recipes, with a built-in data organizer, and data browsing and
plotting facilities.

Although designed for the reduction of the VIMOS VLT Deep Survey 
(VVDS) data, VIPGI is sufficiently general to make 
it able to handle all VIMOS available observing modes and instrument
configurations. 
Its capabilities and the quality of its data reduction pipeline have 
been verified through the reduction of some 30,000 spectra from the
VVDS, demonstrating it can
be used to achieve a data reduction accuracy comparable to the one
obtained using standard IRAF tasks. 
Starting November 2003, VIPGI has been offered for usage to the
astronomical community, but
the scientific validation
process has not, so far, covered all VIMOS observing modes, and all
VIMOS grisms. Therefore at the moment we require it to be used either
in Milano or in Marseille, under the supervision of a VIRMOS
Consortium astronomer, who can provide guidance on how to handle non
standard situations.

\section{The Vimos Interactive Pipeline Graphical Interface}

\subsection{The instrument model}
VIPGI reduction recipes are based onto an instrument model,
which analytically describes the main calibration relations 
required for spectra extraction. The instrument model is separated
into three different components:
the optical distorsion model 
(ODM), is used to convert the positions of an 
observed object on the instrument focal plane to the corresponding 
position on the CCD frame. The curvature model (CM)  
is used to trace the edges of the spectrum along the dispersion 
direction. Finally,  the inverse dispersion 
solution (IDS) is the usual conversion relation 
between the position in pixels on the CCD and the wavelength coordinate.
All these models have been implemented into 
the code as polynomial relations. 
Part of the calibration (like e.g. grism characteristics)
is provided by tables appended to 
raw files during the data organization step,
while another part (namely first guesses for each polynomial 
coefficient) is contained into raw files headers. 
The pipeline uses these first guesses as a starting point 
for the detailed determination of the polynomials.

\subsection{The Graphical Interface}

Although VIPGI recipes automate to a very large extent the task of
reducing VIMOS data, they do not address at all two important and
problematic areas of the global data reduction activity. The first one
is that of organizing the large volume of VIMOS data in such a way
that the correct input is always given to the various recipes, since
the recipes do not perform any extensive validation check on their
input data. The second one is that of a quick and easy browsing of the
data at the various stages of data reduction. It is mainly to address
these to points that we have designed VIPGI, the graphical user
interface for the VIMOS data reduction pipeline. Such GUI is written
almost entirely in Python, to simplify the handling of graphical
elements and to take advantage of the object-oriented programming
capabilities offered by that language, with only some auxiliary
functions devoted to the handling of FITS files carried out by a set
of C programs. 

\subsection{Data Organization}

To be used in VIPGI, raw and reduced VIMOS FITS files must be put
under the control of VIPGI data organizer. Files are phisically
organized into a pre-defined directories structure and renamed
following a scheme that extremely simplifies the task of understanding
what the content of each file is. Moreover the data organizer provides
a logical classification of each file, dividing data according to
observation date, instrument mode (MOS, IFU, or Imaging), quadrant,
grism or filter used during the observation, type of observation
(science exposure, flat field or wavelength calibration lamp, bias
frame, and so on), and observation target. These logical categories
are then used to simplify the browsing of data, allowing the user to
select in a very simple way a specific data category of interest, and
to ensure that correct input data is provided at all times to the
VIPGI reduction recipes. 

\subsection{The data reduction pipeline}

VIPGI data reduction is organized in recipes; each one is a
stand-alone 
program that performs one or more reduction steps. 
All VIPGI data reduction recipes are written entirely in C, to ensure
a satisfactory execution speed. 
When designing recipes, we have tried to group together steps which
are 
normally always executed in the same sequence: for 
example, bias subtraction, flat field correction and cosmic ray
cleaning 
in the imaging case have been grouped into one recipe. 
However there is no single "do it all" recipe that can be fed with a
bunch 
of raw data frames to produce completely reduced 
images or spectra, as we assume that the astronomers will need and 
want to check at least some of the intermediate data 
reduction steps. To help astronomers keep the details of the data 
reduction process under control, the detailed behavior of 
each recipe can be customized via a set of input parameters, 
that are stored in a parameter file.\\

The building blocks for the recipes code are provided by 
a set of routines and library functions
designed to handle VIMOS data. The routines range from the basic
"opening a file and reading its content" to spectral extraction,
wavelength calibration and 3D data cube reconstruction. General tasks
like the handling of FITS files, or of the World Coordinate System,
and the detection of stars within imaging
exposures (for photometric or astrometric calibration purposes) are
carried out by special purpose external software packages (the CFITSIO
and WCSTools libraries, and SExtractor, respectively) that have been
included within the DRS.

The recipe input
and output are always in the form of FITS files, and to avoid
increasing the already large number of files produced by VIMOS, the
different reduction mid-products (like e.g. 2D extracted,
sky-subtracted spectra) together with the
various calibration tables useful for the reduction are "appended" as
extensions to the original FITS files, instead of creating independent
ones. This is done not to save disk space (a multiple extensions file
occupies the same amount of bytes as the sum of many single files),
but to simplify the astronomer's work, as the data reduction process
would otherwise lead to the creation of thousands of files.

For a detailed description of the reduction recipes, we refer to
\cite{Franz04} and \cite{Zanic04}

\subsection{Quality checks, plotting and analysis}

When starting the reduction of an observation run, the first 
step to be performed is usually the refinement of the
first guesses written into the headers. Small variations in the 
instrument observing conditions might results in variations 
of the ODM, CM and IDS coefficients; in this case 
the first guesses might be not accurate enoguh to let the recipes 
find and reduce spectra. Graphical tools are provided within VIPGI to 
inspect and eventually adjust first guesses.

Once first guesses are reasonable for the data at hand, we can proceed
with locating spectra on the CCD.
A good location of spectra is fundamental to not to increase the noise
during the spectral extraction phase. 
In general, spectra are located within one pixel from the expected position, 
provided that, for MOS observations, 
the mask layout designed automatically by the VIMOS Mask Preparation 
Software (VMMPS, \cite{vmmps}) is not altered 
by the astronomer by placing manually slits on the mask,
thus risking to get spectra which overlap on the CCD. 
However, instrument flexures and/or badly postioned masks can differentially displace spectra
from the theoretical position of a relatively large amount (up to 10
pixels), therefore some check on this step is crucial.
To visually check the quality of spectra location, it is possible 
to display an image where the "edges" of the spectra for each 
MOS slit or IFU fiber are superimposed. 

Once spectra are located, the Inverse Dispersion Solution is computed 
for each spectrum. 
The accuracy of the wavelength calibration changes sligthly from 
grism to grism, but the rms residuals around the best fitting 
relation typically amount to better than one fifth of a pixel. 
To visually inspect the goodness of the wavelength calibration, 
a few plots are available,
like e.g. the plot of a single arc line residuals from the best fitting 
solution for each slit in a MOS mask or each fiber in an 
IFU observation and an histogram 
showing the distribution of rms residual values for all slits in a 
MOS or all fibers in an IFU exposure.
As an example, in figure 1 the residuals rms 
histogram plot for an IFU arc lamp exposure with the 
``High Resolution Orange'' grism is shown.
\begin{figure*}[t!]
\centerline{\psfig{figure=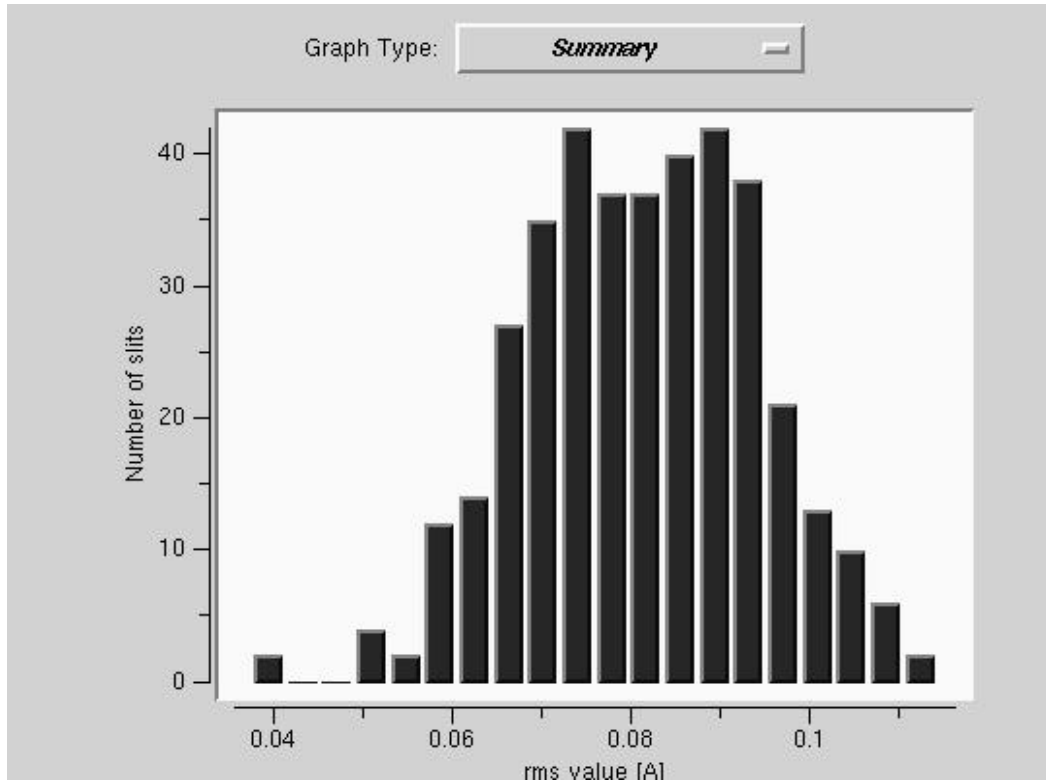,width=14cm}}
\caption{Residuals rms 
histogram plot for an IFU arc lamp exposure}
\label{lambda}
\end{figure*}

A number of different possibilities exist for browsing through 
one-dimensionally and two-dimensionally extracted spectra. 
Two-dimensional slit spectra for a MOS jitter sequence can be 
plotted together with all the single-exposures 2D spectra
used for the combination, to check the reality of spectral 
features and the quality of fringing and sky residuals
removal.

A tool for plotting and analyzing the extracted one-dimensional 
spectra is also provided. This tool allows the astronomer to 
plot each of the one-dimensionally extracted spectra, 
together with the corresponding two-dimensional and sky 1D spectra. 
In this way it is possible to check on the reality of spectral 
features that are present in the one-dimensional spectrum, 
which could be due to sky subtraction or fringing residuals. 
Basic plotting interactive tools, like zooming, smoothing, and 
line fitting are also provided.
The astronomer can obtain quick redshift estimates by 
fitting or marking the position of a set of spectral lines, and 
using a function  that will compute a list of possible redshifts 
based on a list of known lines in galaxy spectra. 
Once the user has chosen one possible solution, the expected 
positions of other lines in the list are marked on the plot,
to visually inspect the goodness of the redshift determination.

A ``summary'' plotting tool is also provided, which incorporates 
in one display window the functionality of the one-dimensional 
spectrum, of the lambda calibration, and of the cross-dispersion 
slit profile plotting tools. All functionalities from the 
one-dimensional plotting tool are preserved. In addition, it is 
also possible to display information on the astronomical 
object whose spectrum is being plotted (see figure 2).

\begin{figure*}[t!]
\centerline{\psfig{figure=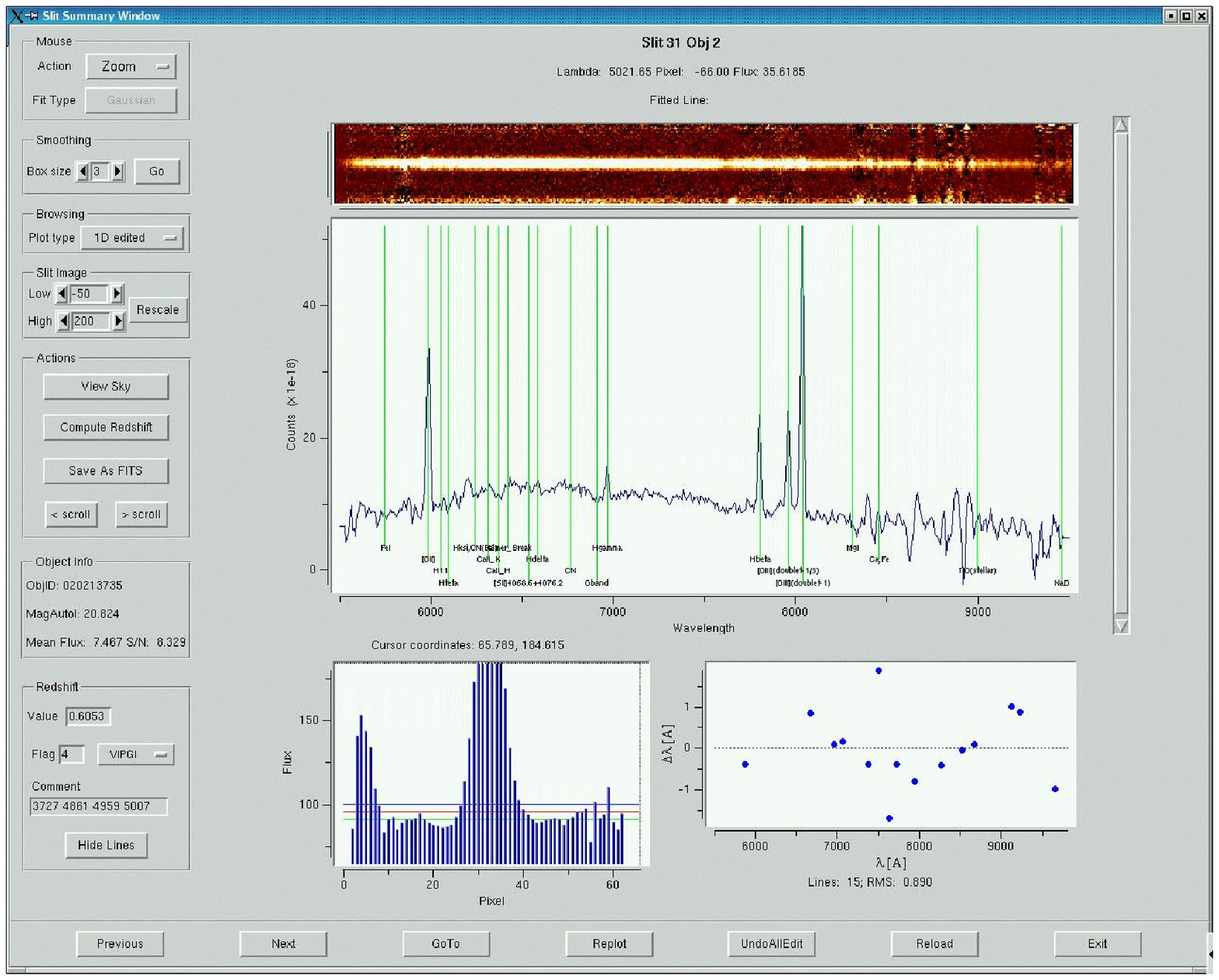,width=18cm}}
\caption{The summary plotting tool: from top to bottom, the 2D
  spectrum, the 1D spectrum, the cross-dispersion 
slit profile(left) and the wavelength calibration rms for the slit
  (right). On the left column, the funtion buttons and some additional objects information.}
\label{lambda}
\end{figure*}

\subsection{Overall quality and performances}

To test the quality of the results from our pipeline we have performed parallel reduction of 
same data with VIPGI and IRAF. This operation
has shown that the quality of our spectra, in terms of continuum 
shape and signal to noise ratio, is basically the same as the 
one obtained with a manual IRAF reduction.
We also have checked our data quality by comparing spectra of 
objects observed twice during VVDS runs. Checking almost two 
hundreds spectra we have determined that the global incertainty 
in redshift measurements on spectra reduced by our
pipeline is of the order of $\Delta$z $\sim$ 0.001 (see \cite{LefevreCDFS}). 
An example of extracted 1D spectrum is shown in figure 3
\begin{figure*}[t!]
\centerline{\psfig{figure=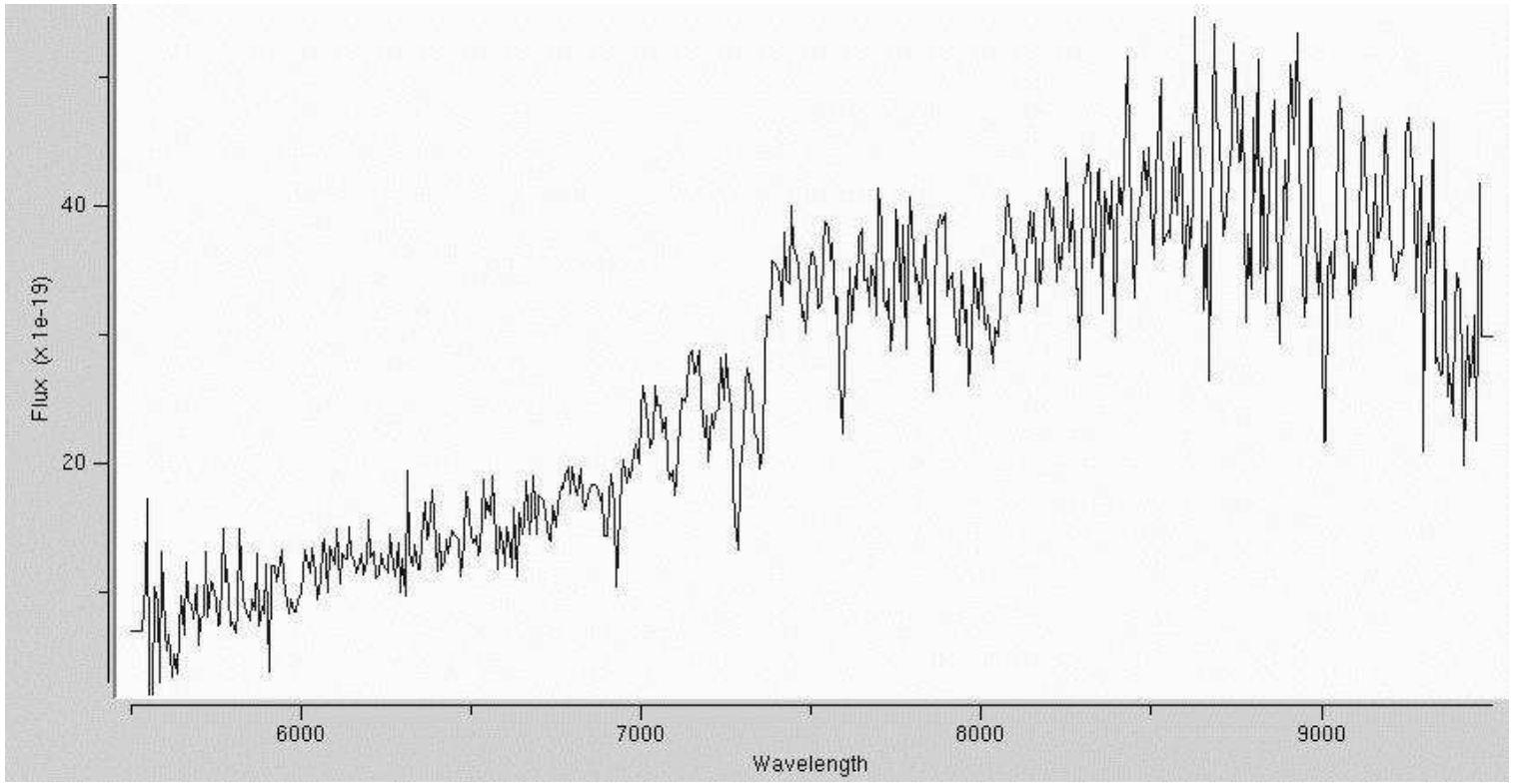,width=20cm}}
\caption{1D spectrum of a VVDS galaxy at z=0.851. The Ca H and K lines
and the Balmer break are clearly visible}
\label{spec}
\end{figure*}

A complete reduction for one quadrant of a single VIMOS MOS or 
IFU pointing, going from the raw data to the derivation of 
the wavelength calibration, and to the extraction of one-dimensional 
spectra, flux and wavelength calibrated and sky subtracted, 
takes typically around 5 minutes on a Linux PC equipped with an 
AMD Athlon 2200+ processor and 1 GB of RAM. 
On the same PC a full reduction for all 4 quadrants of a jitter 
sequence of 10 MOS or IFU observations can be carried out in 
approximately 30 minutes (not counting the time eventually 
dedicated to the visual checks of the data reduction 
results). The main factor influencing the speed of the data 
reduction process is the amount of physical memory (RAM) available
to the pipeline. 
On computers with less than 512 MB of RAM VIPGI 
recipes would require a much longer time to execute.\\

\section {Present and future developments}

After one year of extensive usage, we can safely state that VIPGI has
received an excellent feedback from the users, the most
appreciated features being the capabilities of data organization, and
the possibility to carry out all the reduction AND the first analysis
steps within the same environment.
On the other hand, the basic limitation of VIPGI is that of being
``VIMOS centric'': the Data Organizer, but also the plotting and data
inspection facilities outlined above, have been deisgned and developed 
having VIMOS in mind, and it is extremely difficult, if not
impossible, to export these tools to be used with other kinds of data.

However, the underlying concept, i.e. an environment allowing Data Organization 
and handling communications between different analyisis, reduction and 
display tools, can indeed be exported and developed so that it can be
used for any kind of astronomical data, at least in the domain of
optical and Near Infrared instruments.
This is the basic idea of our new project DRAGO 
(Data Reduction and Analysis Graphical Organizer).
DRAGO has the ambition of providing the user with an easily configurable
environment, within which different tools defined by the user can be
plugged in.

DRAGO is structured around a powerful data organizer, which
provides the user with the facility to select in very simple ways
(sub)sets of the available data to work with. The storage and
selection functions for the data organizer are handled by either a
database server embedded within the package itself, or an external
database and/or data archive server, in case the user has already
stored and organized his/her data  in a database of choice, or is
using data from a public data archive. Working in connection with the
data organizer there is a series of operational modules that provide
the interface to the data reduction and analysis tools. Examples of
such modules are the VIMOS data reduction pipeline, the 1-dimensional
spectra visualization tool, the  Euro3D IFU datacube visualization
tool, redshift and spectral line parameter measurement tools. 

To fully take advantage of the package functionalities users will have
to "import" data into the system, and classify them according to a set
of user-defined classification rules, or provide access to already
existing database tables. 

DRAGO is currently under development, and we refer to Paioro et al,
these same proceedings, for a more detailed explanation of
its structure and its status

\section{acknowledgements}
We thank all the VIRMOS consortium for having extensively tested and 
used VIPGI for reduction, giving us unvaluable suggestions
for improvement: we especially thank S.Paltani for pointing out
bugs and helping us in their solvimng, and E.Zucca for
her precise and puntilious comments.
We also thank R.Palsa P. Sartoretti and C.Izzo, from ESO DMD division, for
the fruitfull collaboration in developing the DRS basic functions.
This work has been partially supported by CNR, COFIN2000, COFIN2003  and INAF.

\end{document}